\documentclass[prl,aps,reprint,twocolumn,showpacs,superscriptaddress]{revtex4-2}
\usepackage{physics}
\usepackage[english]{babel}
\usepackage[utf8]{inputenc}
\usepackage[pdftex]{graphicx}
\usepackage{dsfont}
\usepackage{bm}
\usepackage[normalem]{ulem}
\usepackage{hyperref}
\usepackage{epstopdf}
\usepackage{color}
\usepackage[sort&compress]{natbib}
\usepackage{subfigure}
\usepackage{colordvi}
\usepackage{epsfig}
\usepackage{float}
\usepackage{dcolumn}
\usepackage{ifpdf}
\usepackage{graphicx,color,bm}
\usepackage[english]{babel}
\usepackage{amsmath,amssymb}
\usepackage{stmaryrd,dsfont,txfonts}
\usepackage{subfigure}
\usepackage{hyperref}
\hypersetup{
	colorlinks,%
	citecolor=blue,%
	linkcolor=blue,%
	urlcolor=blue
}

\begin{document}
\title{Strain-Driven ``Sinusoidal'' Valley Control of Hybridized $\Gamma -\mathrm{K}$ Excitons }

	\author{Yingtong Zhu}
	\affiliation{Department of Physics, Qufu Normal University,  Qufu 273165, Shandong, China}
        \author{Kang Lan}
	\affiliation{Department of Physics, Qufu Normal University,  Qufu 273165, Shandong, China}
        \author{Shiling Li}
	\affiliation{Department of Physics, Qufu Normal University,  Qufu 273165, Shandong, China}
                                  \author{Ning Hao}
                        \affiliation{Anhui Province Key Laboratory of Low-Energy Quantum Materials and Devices , High Magnetic Field Laboratory, HFIPS, Chinese Academy of Sciences, Hefei, Anhui 230031, China}
      \author{Ping Zhang}
\affiliation{Department of Physics, Qufu Normal University,  Qufu 273165, Shandong, China}
\affiliation{Beijing Computational Science Research Center, Beijing 100084, China}
	         \author{Jiyong Fu}
	\thanks{yongjf@qfnu.edu.cn}
	\affiliation{Department of Physics, Qufu Normal University, Qufu 273165, Shandong, China}
        \begin{abstract}
          The photoluminescence (PL) of momentum-indirect $\rm \Gamma- K$ excitons in monolayer WS$_2$ under biaxial strain was recently observed by Blundo \emph{et al.} [Phys. Rev. Lett. \textbf{129}, 067402 (2022)], yet its microscopic origin remains elusive.
          Here we develop a unified framework that reproduces the measured  PL and reveals its fundamental excitonic mechanism.
          We reveal that: (i) the PL originates from genuinely  hybridized direct-indirect excitonic \emph{eigenstates}, rather than \emph{nominally} mixed species with
          fixed dominant character; (ii) the direct exciton converts into the indirect one via a previously unrecognized two-step pathway---exchange-interaction-driven
          exciton transfer followed by a spin flip; and (iii) a higher-energy indirect exciton, absent from prior studies, acts as a crucial \emph{intermediate} mediating this conversion.         Beyond explaining experiment, our theory predicts a striking strain-driven {``sinusoidal''} valley response, furnishing a continuously tunable \emph{valley dial} that far exceeds binary control schemes. This unified picture of strain-engineered direct-indirect exciton dynamics introduces a new paradigm for manipulating long-lived valley degrees of freedom, opening a pathway toward programmable valley pseudospin engineering and next-generation valleytronic quantum technologies.
                    \end{abstract}
\maketitle
\paragraph*{Introduction.---}Monolayer transition-metal dichalcogenides (TMDCs) have emerged as  a versatile platform for exploring excitonic and valleytronic phenomena in two dimensions~\cite{BSE1,TMDc1}. Owing to their strong Coulomb interactions~\cite{Coulombinteractions1,Coulombinteractions2} and reduced dielectric screening, these materials host tightly bound excitons in diverse configurations, including momentum-direct (bright or spin-forbidden dark)~\cite{brightexciton1,brightexciton2,darkexciton2,darkexciton3,darkexciton1,darkexciton4} and indirect (momentum-forbidden dark)~\cite{PRL,StrainQ,FariaJunior2022} states.
The coexistence and coupling of these direct and indirect species offer a unique window into spin-valley interactions.
Understanding how these states evolve under external engineering knobs---such as strain~\cite{Kumar2024,HenriquezGuerra2023,Stellino2024,FariaJunior2022}---is essential for tailoring light-matter interactions and exciton dynamics in TMDC-based valleytronic devices.

Recent experiments, supported by the first-principles calculations~\cite{PRL}  have  uncovered  clear photoluminescence (PL) signatures  of indirect $\Gamma-\mathrm{K}$ excitons in monolayer WS$_2$. However,  interpretation based on \emph{nominally}  mixed direct  or indirect excitons,  where each state retains one fixed dominant component, fails to capture the observed peak behavior or its  strain dependence. Moreover, the microscopic pathway that transfers population between direct and indirect species remains unresolved. Given the inherently long lifetime of indirect  excitons and their potential application for coherent quantum control, a unified theoretical model incorporating  strain-mediated PL, valley polarization (VP), and the full set of intra- and intervalley relaxation channels is highly desirable---for both fundamental understanding and the realization of strain-programmable valley pseudospin functionalities.

\begin{figure}[h]
	\includegraphics[width=7.5cm]{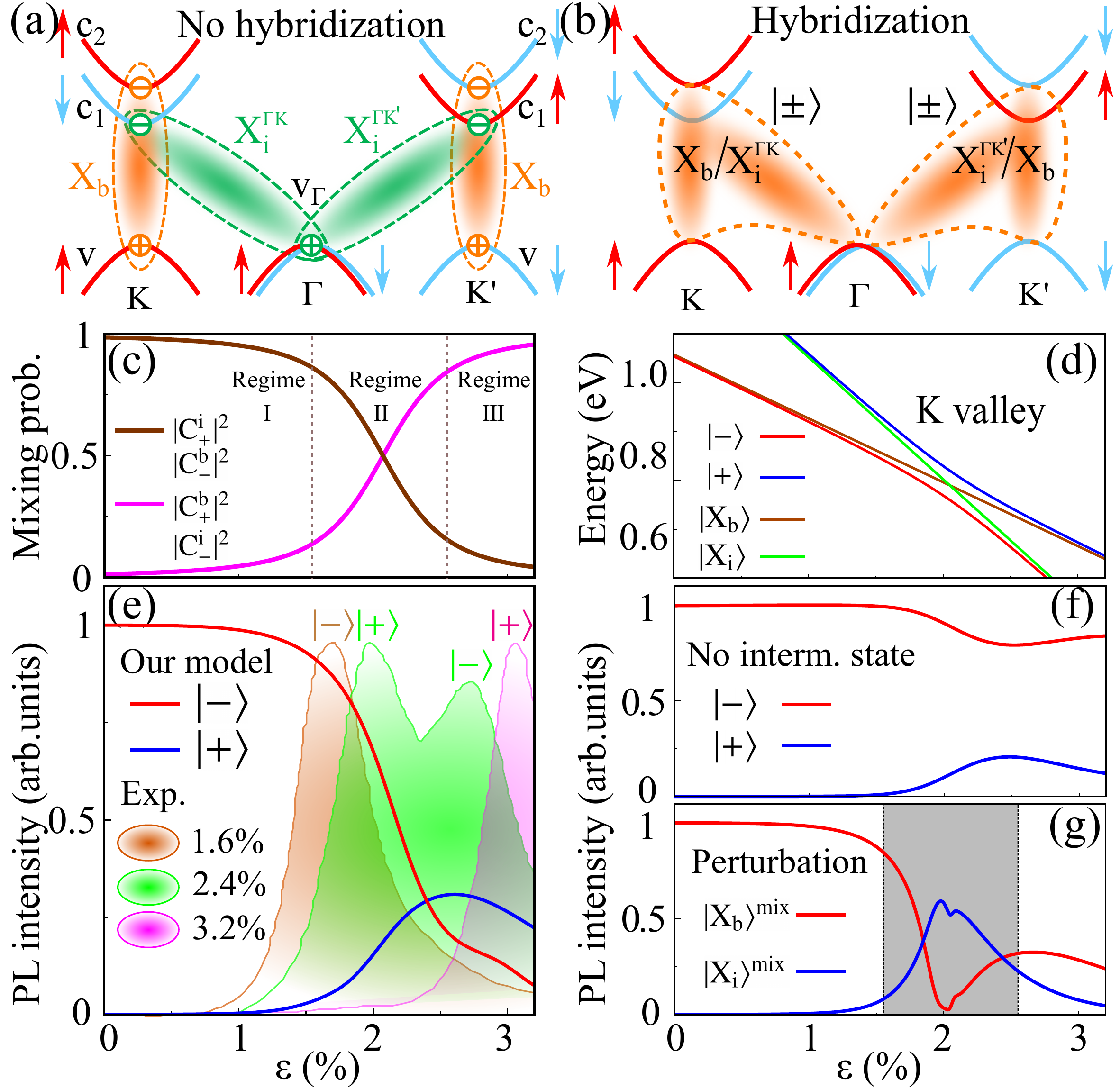}
	\centering
	\caption{(Color online)  (a) Spin-valley configurations of the direct exciton ($\mathrm{X_b}$) and momentum-indirect excitons ($\mathrm{X_i}^{\Gamma\mathrm{K}}$, $\mathrm{X_i}^{\Gamma\mathrm{K'}}$), involving the spin-split CB branches ($c_1, c_2$), the upper VB ($v$), and the $\Gamma$-valley VB ($v_\Gamma$).
(b) Schematic hybridization between $\mathrm{X_b}$ and $\mathrm{X_i}$ forming the eigenstates $|\pm\rangle$.
          (c) Strain dependence of  mixing coefficients $C_\pm^{\mathrm b}$ and $C_\pm^{\mathrm i}$, with the dominant component of $|\pm\rangle$ interchanging near
          $\varepsilon \approx 2\%$. (d) Strain evolution of excitonic energies.
          (e–g) PL intensity vs strain: (e) full model (red/blue) compared with experimental data [shaded (purple, green, pink)  regions] from Ref.~\cite{PRL}; (f) model without the intermediate state; (g) perturbative approach, with the shaded (gray) region [referring to Regime II in (c)] marking where the perturbation breaks down. }
           	\label{FIG. 1.}
\end{figure}

\begin{figure}[h]
	\includegraphics[width=6.8cm]{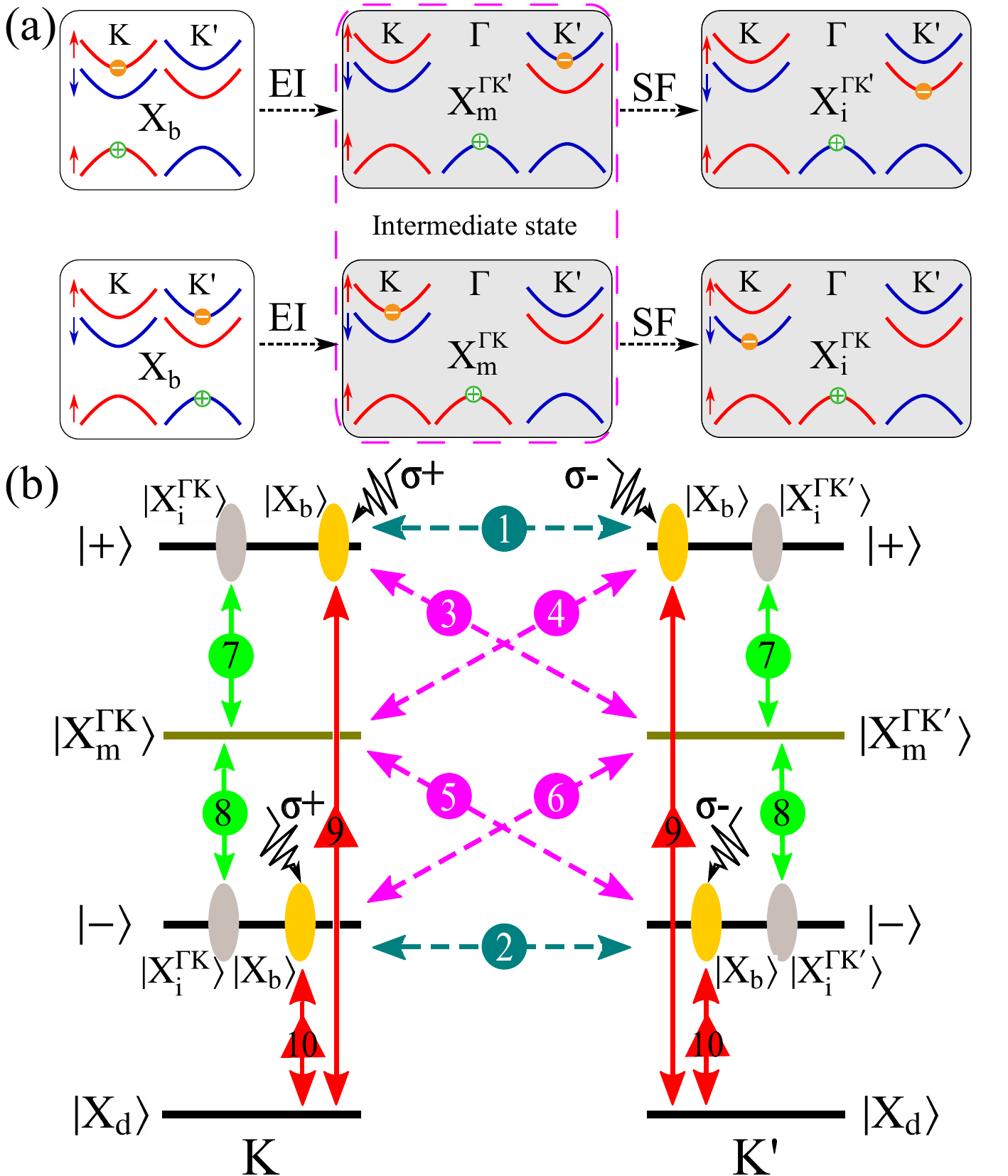}
	\centering
	\caption{(Color online)  (a) Upper: Transition pathway from the direct exciton $\mathrm{X_b}$ in the K valley to the indirect exciton $\mathrm{X_i^{\Gamma K'}}$ in the K$^\prime$ valley, mediated by exchange interaction~\cite{footnote-exchange} and  spin flip, with $\mathrm{X_{m}^{\Gamma K'}}$ as the intermediate state. Lower: time-reversed counterpart of the upper panel.
          (b) Intra- (solid arrows, 7–10) and intervalley (dashed arrows, 1–6) relaxation channels among hybridized ($|+\rangle$, $|-\rangle$) and non-hybridized ($\mathrm{X_m}$, $\mathrm{X_d}$) excitonic states under linearly polarized excitation. The hybridized states contain direct (yellow) and indirect (gray) components. Double arrows denote reversible up- and down-conversion processes, with the latter energetically favored.
                  }
 	\label{FIG. 2.}
\end{figure}

\begin{figure}[h]
	\includegraphics[width=6.5cm]{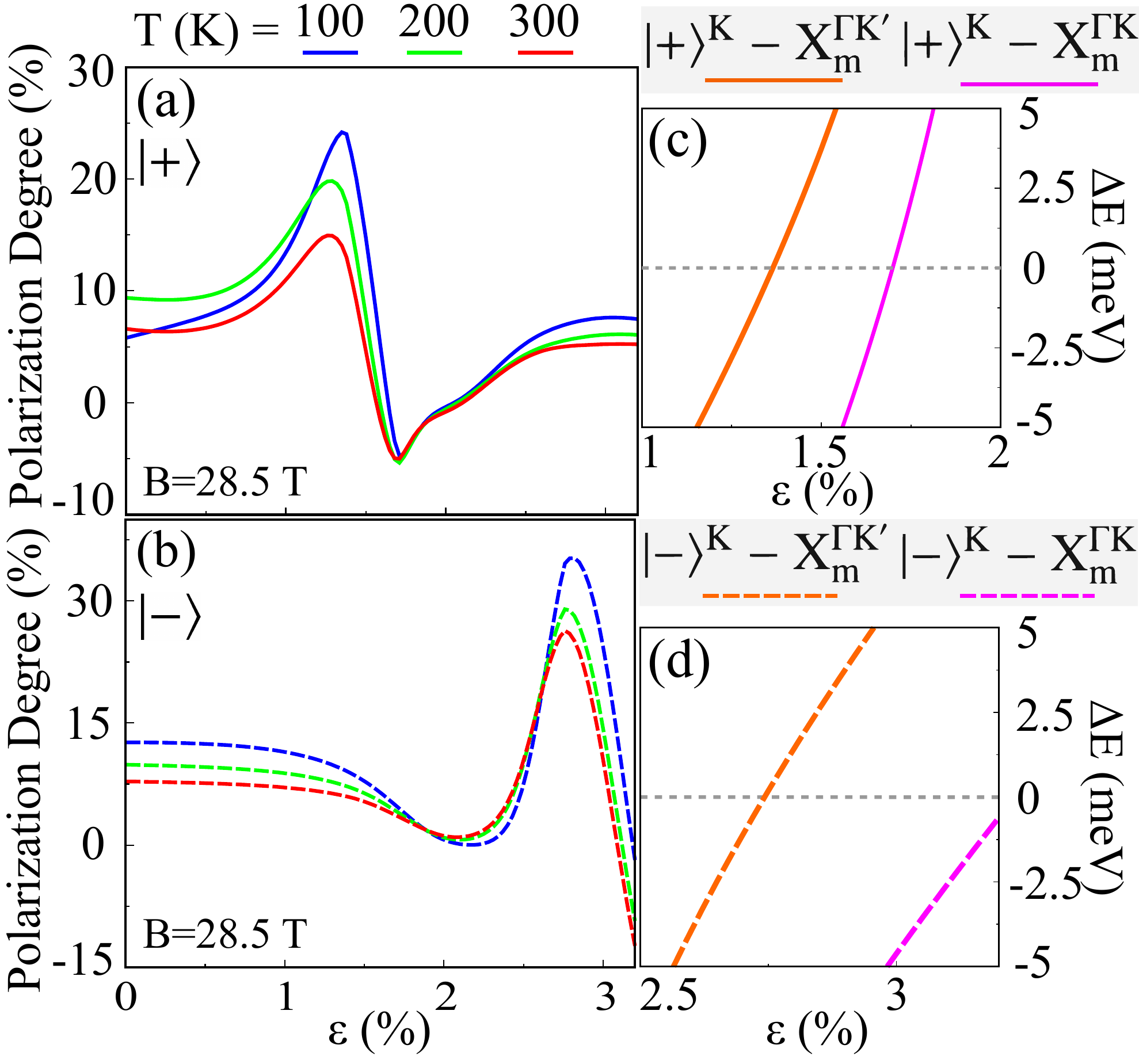}
	\centering
	\caption{(Color online) (a), (b) Valley polarization of $|\pm\rangle$ versus strain at $T=100$, 200, 300 K.
          (c), (d) Energy separation $\Delta E_{\pm, m}^{\tau}=E_{|\pm\rangle}^{\rm K}-E_{\rm X_m}^{\rm \tau}$ between  $|\pm\rangle$  in the K valley and
          the intermediate state $\rm X_m^{\Gamma K}$ (K valley) and $\rm X_m^{\Gamma K^\prime}$ (K$^\prime$ valley) as functions of  strain.
                            }
 	\label{FIG. 3.}
\end{figure}

Here, we bridge this gap by developing a unified framework that captures the strain, thermal, and magnetic responses of both direct and indirect excitons,
achieving excellent  agreement with experiment [Fig.~\ref{FIG. 1.}(e)]. We show that the observed PL originates from hybridized excitonic \emph{eigenstates} rather than \emph{nominally} mixed species. We further identify a two-step conversion mechanism---exchange interaction followed by spin flip---and uncover a previously unrecognized higher-energy state that acts as a crucial \emph{intermediate} mediating this conversion [Fig.~\ref{FIG. 2.}]. Strikingly, our model predicts a robust  {``sinusoidal’''} VP control in the upper hybridized branch [Fig.~\ref{FIG. 3.}(a)], enabling a continuously tunable \emph{valley dial} far more versatile than binary control schemes.
These results establish a unified picture of strain-engineered  valley physics, and introduce a new concept:  ``sinusoidal'' valley control, opening a pathway for
leveraging  long-lived indirect excitons in valley  information processing, logic operations, and next-generation valleytronic architectures.

\paragraph*{Model Hamiltonian.---} The spin-orbit coupling (SOC) endows TMDCs with  spin-valley \emph{locked} band structure at the K and K$^{\prime}$ valleys of the Brillouin zone~\cite{valleyHalleffects,spinvalleylocking,soc1,soc2}. The upper valence band (VB;$v$) and two spin branches of the conduction band (CB)---lower ($c_1$) and upper ($c_2$)---constitute the basis for direct excitons [Fig.~\ref{FIG. 1.}(a)]:  bright ($\mathrm{X_b}$) and dark ($\mathrm{X_d}$; spin-forbidden) excitons. With increasing strain, while the CB minimum remains at the K valley, the VB maximum shifts toward the $\Gamma$ point ($v_{\Gamma}$), driving a crossover from direct to indirect band gap~\cite{FariaJunior2022,evidence}.
 This evolution favors the conversion  of direct ($\mathrm{X_b}$) into
 indirect ($\mathrm{X_i}$) excitons, where  $\mathrm{X_i^{\Gamma K}}$ ($\mathrm{X_i^{\Gamma K^{\prime}}}$) consists of an electron in the  K (K$^\prime$) valley ($c_1$)  bound to  a hole in the $\Gamma$ valley [Fig.~\ref{FIG. 1.}(a)].
 We further introduce a previously unrecognized higher-energy indirect exciton $\mathrm{X_{m}^{\Gamma\mathrm{K}}}$ ($\mathrm{X_{m}^{\Gamma K^{\prime}}}$) [Fig.~\ref{FIG. 2.}(a)] with  electron residing  in upper $c_2$  band.
    
 Symmetry analysis reveals that the direct-indirect excitonic hybridization between $\mathrm{X_b}$ and  $\mathrm{X_i}$ occurs~\cite{PRL},  through their shared electronic states residing in the same valley [Fig.~\ref{FIG. 1.}(b)]. The resulting multiexcitonic Hamiltonian under the basis  $\big \{ |\mathrm{X_b^{\rm K}}\rangle, \mathrm{|X_i^{\rm \Gamma K}}\rangle,  \mathrm{|X_b^{\rm K^\prime}}\rangle, \mathrm{|X_i^{\rm \Gamma K^\prime}}\rangle \big\}$  reads~\cite{PRL}, 
 \begin{eqnarray}
   \nonumber \mathcal{H} = \mathcal{E}_+\varsigma_0\otimes \tau_0&-&\mathcal{E}_-\varsigma_z\otimes \tau_0+\Delta \mathcal{E}_+\varsigma_0\otimes \tau_z-\notag\\
   \Delta \mathcal{E}_-\varsigma_z\otimes\tau_z
    &+&\Delta_R\varsigma_x \otimes \tau_0+\Delta_I\varsigma_y\otimes \tau_z,
   \label{Eq:H_compact}
   \end{eqnarray}
 where $\varsigma_0$ ($\tau_0$) is the identity matrix and $\varsigma_{x,y,z}$ ($\tau_{x,y,z}$) represents  the Pauli (``pseudospin'') matrices acting within the
 exciton (valley) subspace. We define $\mathcal{E}_\pm=(E_{\rm X_b}^0\pm E_{\rm X_i}^0)/2$ and $\Delta \mathcal{E}_\pm=(\Delta E_{\rm X_b}\pm \Delta E_{\rm X_i})/2$,
 in which  $E_{j}^0$ is the zero-field ($B=0$) energy of ${\rm X}_j$  and $\Delta E_{j}=(1/2)g_j\mu_B B$ represents valley-Zeeman shift, with
 $g_j$ the $g$ factor ($j=\rm X_b, X_i$). The parameter $\Delta=\Delta_R + i \Delta_I$ characterizes the hybridization (coupling)  between   $\rm X_b$  and $\rm X_i$ with  $\Delta_R$ ($\Delta_I$) the real (imaginary) components.

 \paragraph*{Multi-excitonic energies with strain and magnetic responses.---}We adopt the the two-band  model proposed by Xiao \emph{et al.}~\cite{spinvalleylocking} to determine the single-particle band edges. We then incorporate many-body interactions to obtain the excitonic binding energies by solving the Bethe-Salpeter equation (BSE)~\cite{Bz1,BSE1,BSE2,BSE3,BSE4,BSE5}, where the exciton state is expressed as $\Psi_n = \sum_{c,v,\mathbf{k}} J_{c,v,\mathbf{k}}^{(n)} |v\mathbf{k} \rightarrow c\mathbf{k}\rangle$, with $c$ ($v$) denoting the conduction (valence) band, $\mathbf{k}$ the wave vector, and $J$ the BSE expansion coefficient. To further refine the single-particle states for evaluating many-body corrections, we employ an 11-band tight-binding model
 that includes second-nearest-neighbor hopping terms~\cite{11band,11band1}.
With these considerations, we obtain the zero-$B$-field energy of $\rm X_b$ and  $\rm X_i$,
\begin{align}\label{EXB0}
  E_{\rm X_b}^{0}=E_g+\frac{1}{2}\Delta_c-\frac{1}{2}\Delta_v-V_{\rm X_b}+\delta_{\rm X_b}{\varepsilon}, \\
  \label{EXI0}
  E_{\rm X_i}^0=E_g+\frac{1}{2}\Delta_c-\Delta_\Gamma-V_{\rm X_i}+\delta_{\rm X_i}\varepsilon,
\end{align}
where $\Delta_{\Gamma}$ is the $\rm \Gamma-K$ VB offset.  We consider the exciton binding energies as $V_{\rm X_b}=V_{\rm X_i}=\langle V_{e-h}\rangle$, the expectation value of the electron-hole Coulomb interaction from BSE. The last term in Eq.~\eqref{EXB0} [Eq.~\eqref{EXI0}] describes the strain-induced energy shift, which is approximately linear in the considered strain range~\cite{PRL}, with the  strain response coefficients $\delta_{\rm X_b}$ [$\delta_{\rm X_i}$].

To determine valley-Zeeman shifts, rather than invoking the empirical decomposition into spin, valley and
atomic-orbital contributions~\cite{opticalselectioni}, we treat valley and orbital components on equal footing via the  orbital angular momentum $\mathbf{L}$, evaluated from first-principles calculations~\cite{gfactor,g1,g2,g3}. Namely, for a Bloch state $|n\textbf{k}\rangle$  the  effective $g$ factor is $g_{n,\textbf{k}}=L_{n,\textbf{k}}+S_{n,\textbf{k}}$, giving the $\tau$-valley exciton energies under $B$ field $E_{j}^{\tau}(B)=E_{j}^{0}+\tau \Delta E_j(B)$, $j=\rm X_b, X_i$.
For  other excitonic energies, including the spin-forbidden dark  ($\rm X_d$) and the higher-energy indirect ($\rm X_{m}$) excitons [Fig.~\ref{FIG. 2.}(a)], see the Supplementary Material (SM). Note that, for notational simplicity, here the valley index of indirect exciton  $\rm X_{i(m)}^{\mathrm{\Gamma K(\Gamma K^\prime)}}$ is  assigned according to the electron constituent only, i.e., $\rm X_{i(m)}^{\mathrm{\Gamma K(\Gamma K^\prime)}}\equiv \rm X_{i(m)}^{\mathrm{K(K^\prime)}}$.

\paragraph*{Hybridized direct-indirect excitonic eigenstates and nominally mixed excitons.---}From Eq.~\eqref{Eq:H_compact}, the hybridized ``$\mathrm{X_b/X_i}$''
eigenstates ($|\pm\rangle$) in the K valley and the corresponding eigenenergies ($E_{|\pm\rangle}$) are obtained as
$|\pm\rangle = C_{\pm}^{\rm b}|\mathrm{X_b}\rangle + C_{\pm}^{\rm i}|\mathrm{X_i}\rangle$ and
$E_{|\pm\rangle} = E_+ \pm \sqrt{E_-^2 + |\Delta|^2}$,
where $E_{\pm} = (E_{\mathrm{X_b}}^{\rm K} \pm E_{\mathrm{X_i}}^{\rm K})/2$.
Clearly, the mixing coefficients depend on the relative energy ordering of $\mathrm{X_b}$ and $\mathrm{X_i}$, which coincide near $\varepsilon\sim 2.0\%$ [Fig.~\ref{FIG. 1.}(d)].  For $E_{\mathrm{X_b}} < E_{\mathrm{X_i}}$ [Fig.~\ref{FIG. 1.}(c)],  they satisfy
$|C_{+}^{\rm b}|^2 = |C_{-}^{\rm i}|^2 = \sin^2(\theta/2)$ and $|C_{-}^{\rm b}|^2 = |C_{+}^{\rm i}|^2 = \cos^2(\theta/2)$, with $\tan\theta = |\Delta|/E_-$.
Particularly, in Regime I [Fig.~\ref{FIG. 1.}(c); $\varepsilon<1.55$], the lower ($|-\rangle$) and upper ($|+\rangle$) branches are predominantly $\mathrm{X_i}$-like and $\mathrm{X_b}$-like, respectively [cf. Figs.~\ref{FIG. 1.}(c) and \ref{FIG. 1.}(d)].
When the energy ordering reverses ($E_{\mathrm{X_b}} > E_{\mathrm{X_i}}$), the dominant characters of the two hybridized branches interchange---namely, the lower (upper) branch becomes $\mathrm{X_b}$-like ($\mathrm{X_i}$-like) in Regime III.
Near the crossover point (Regime II), corresponding to $\varepsilon \sim 2\%$, the two types of excitons are strongly hybridized,
exhibiting comparable $\mathrm{X_b}$ and $\mathrm{X_i}$ character. These features  are important  for interpreting experimental data in Ref.~\cite{PRL}.

We further determine the \emph{nominally mixed} states using the perturbative approach, expressed as 
$|\mathrm{X_b}\rangle^{\rm mix} = C_{\rm b}^{\rm b}|\mathrm{X_b}\rangle + C_{\rm b}^{\rm i}|\mathrm{X_i}\rangle$ and
$|\mathrm{X_i}\rangle^{\rm mix} = C_{\rm i}^{\rm i}|\mathrm{X_i}\rangle - C_{\rm i}^{\rm b}|\mathrm{X_b}\rangle$,
with $| C_{\rm b}^{\rm b}|^2 = | C_{\rm i}^{\rm i}|^2 = 4/(4 +\tan^2\theta)$ and  $| C_{\rm b}^{\rm i}|^2 = | C_{\rm i}^{\rm b}|^2 = \tan^2\theta/(4 +\tan^2\theta)$.
Although both $|\mathrm{X_b}\rangle^{\mathrm{mix}}$ and $|\mathrm{X_i}\rangle^{\mathrm{mix}}$ contain finite admixtures of $\mathrm{X_b}$ and $\mathrm{X_i}$ components, each remains primarily dominated by its own parent state. This behavior contrasts with that of the genuinely   hybridized {eigenstates} ($|\pm\rangle$), whose dominant character  {continuously} interchanges between $\mathrm{X_b}$ and $\mathrm{X_i}$ with strain, as aforementioned.
As a consequence, far from the avoided crossing, 
$|+\rangle \rightarrow \rm X_i$ and $|-\rangle \rightarrow \rm X_b$ for $\varepsilon < 1.55\%$ [Regime I], and   $|+\rangle \rightarrow \rm X_b$ and $|-\rangle \rightarrow \rm X_i$ for  $\varepsilon < 2.55\%$ [Regime III]. 
For quantitative comparison of the mixing amplitudes $C_{\pm}^{\mathrm{b(i)}}$ (for $|\pm\rangle$) and $C_{\mathrm{b(i)}}^{\mathrm{b(i)}}$ (for $|\mathrm{X_{b(i)}}\rangle^{\mathrm{mix}}$), see the SM. 

\paragraph*{Excitonic scattering channels and valley dynamics.---}Figure~\ref{FIG. 2.}(b) shows the intra- and intervalley relaxation channels of both hybridized ($|\pm\rangle$) and nonhybridized ($\mathrm{X_{m}}$, $\mathrm{X_d}$) excitonic states under the linearly polarized excitation, which coherently drives the formation of $\mathrm{X_b}$ in both the K ($\sigma^+$) and K$'$ ($\sigma^-$) valleys.
We reveal that $\mathrm{X_b}$ can convert into the \emph{opposite-valley} indirect exciton $\mathrm{X_i^{\Gamma K'}}$ through two sequential scattering steps [Fig.~\ref{FIG. 2.}(a)]:
(i) exchange-interaction (EI)–induced intervalley scattering to the higher-energy intermediate state $\mathrm{X_{m}}$~\cite{footnote-exchange} [channels 3-6 in Fig.~\ref{FIG. 2.}(b)], followed by
(ii) an intravalley spin-flip (SF) relaxation of $\mathrm{X_{m}}$ into $\mathrm{X_i^{\Gamma K'}}$ (channels 7 and 8).
Simultaneously, $\mathrm{X_b}$ couples to $\mathrm{X_i}$ within the same valley via \emph{shared} electronic states [Fig.~\ref{FIG. 1.}(b) and Eq.~\eqref{Eq:H_compact}], forming hybridized eigenstates $|\pm\rangle$ that comprise both $\mathrm{X_b}$ [yellow frame in Fig.~\ref{FIG. 2.}(b)] and $\mathrm{X_i}$ (gray frame) components.
Additional relaxation pathways include EI-driven intervalley transfer of $\mathrm{X_b}$ (channels 1 and 2) and nonradiative intravalley bright-dark conversion via SF scattering (channels  9 and 10).  Because of the $B$-field-induced lifting of valley degeneracy, we consider  \emph{asymmetric} intervalley transfer of
excitonic states (channels 1-6), with scattering rates~\cite{IntervalleyScattering} ${\gamma_{\pm}=({1}/{\tau_{v0}})[{\Gamma^2}/({\Gamma^2+\Delta E^2})]+{\alpha\Delta E^3}/{|e^{\pm\Delta E/k_BT}-1|}}$,  where $\gamma_-$ ($\gamma_+$) describes transitions from the higher  (lower) to lower  (higher)-energy valley.
The first term represents the electron-hole EI contribution, which acts as an  in-plane field and drives intervalley excitonic transfer with characteristic timescale $\tau_{v0}$ and width parameter $\Gamma$~\cite{IntervalleyScattering}.
The second term accounts for phonon-assisted scattering~\cite{phonon}, governed by the exciton-phonon coupling constant $\alpha$ and the valley Zeeman splitting $\Delta E$. With the above considerations, we obtain the coupled valley dynamic equations,
which for  hybridized states $|\pm\rangle$ in the K valley read,
\begin{eqnarray}
  \frac{dn_+}{dt}=-|C_{\rm +}^{\rm b}|^2 g-\frac{n_{+}}{\tau_+}-\frac{|C_{\rm +}^{\rm b}|^2 n_+}{1/\Gamma_{\rm 9}^{\rm b\Rightarrow d}}+\frac{n_{\rm d}}{1/\Gamma_{\rm 9}^{\rm d\Rightarrow b}}-\frac{|C_{\rm +}^{\rm b}|^2 n_{+}}{1/\gamma_1}\notag\\ +\frac{|C_{\rm +}^{\rm b}|^2 n_{+}^\prime}{1/\gamma_1^{\prime}}-\frac{|C_{\rm +}^{\rm b}|^2 n_{+}}{1/\gamma_3}+\frac{n_{\rm m}^\prime}{1/\gamma_3^{\prime}}-\frac{|C_{\rm +}^{\rm i}|^2 n_+}{1/\Gamma_{\rm 7}^{\rm i\Rightarrow m}}+\frac{n_{\rm m}}{1/\Gamma_{\rm 7}^{\rm m\Rightarrow i}},\\
  \frac{dn_-}{dt}={}-|C_{\rm -}^{\rm b}|^2 g-\frac{n_-}{\tau_-}-\frac{|C_{\rm -}^{\rm b}|^2 n_-}{1/\Gamma_{\rm 10}^{\rm b\Rightarrow d}}+\frac{n_{\rm {d}}}{1/\Gamma_{\rm 10}^{\rm d\Rightarrow b}}-\frac{|C_{\rm -}^{\rm b}|^2 n_-}{1/\gamma_2}\notag\\+{}\frac{|C_{\rm -}^{\rm b}|^2 n_-'}{1/\gamma_{2}^{\prime}}-\frac{|C_{\rm -}^{\rm b}|^2 n_-}{1/\gamma_6}+\frac{n_{\rm {m}}}{1/\gamma_6^{\prime}}-\frac{|C_{\rm -}^{\rm i}|^2 n_-}{1/\Gamma_{\rm 8}^{\rm i\Rightarrow m}}+\frac{n_{{\rm m}}}{1/\Gamma_{\rm 8}^{\rm m\Rightarrow i}},
\end{eqnarray}
where $n_\pm$, $n_{\rm d}$, and $n_{\rm m}$ denote the K-valley densities of $|\pm\rangle$, X$_{\rm d}$, and X$_{\rm m}$, respectively.
And, $\gamma_{1\text{–}6}$ ($\gamma_{1\text{–}6}^\prime$) represent the intervalley scattering rates of excitonic states from the K (K$^\prime$) to the K$^\prime$ (K) valley, with the subscripts indicating the channels 1-6 in Fig.~\ref{FIG. 2.}.
Similarly, $\Gamma_{7\text{–}10}^{\rm \mathrm{A}\Rightarrow \mathrm{B}}$ denote the scattering rates of the intravalley scattering channels 7-10 within the K valley, where $\mathrm{A} \Rightarrow \mathrm{B}$ indicates excitonic transfer from X$_{\mathrm{A}}$ to X$_{\mathrm{B}}$ with $\mathrm{A}(\mathrm{B}) \in {\mathrm{i},\mathrm{m},\mathrm{b},\mathrm{d}}$. For the valley dynamics of  remaining excitonic states and those in the K$^\prime$ valley and relevant parameters used in our calculations, see the SM (Secs.~III and IV).

\paragraph*{Optical activation of hybridized $\rm \Gamma-K$ indirect dark excitons.---}According to experimental reports in Ref.~\cite{PRL} [shaded (colored) regions in Fig.~\ref{FIG. 1.}(e)], the PL response evolves systematically with strain.
Specifically, for $\varepsilon < 1.6\%$ (purple), the PL arises almost exclusively from the direct exciton $\mathrm{X_b}$.
At $\varepsilon = 2.4\%$ (green), both the direct and indirect branches become visible, with $\mathrm{X_b}$ showing slightly stronger emission.
When the strain increases to $3.2\%$ (pink), the PL is dominated by $\mathrm{X_i}$, with the direct excitonic contribution nearly vanishing.
Our theory clarifies that the microscopic origin of the strain evolution arises from the hybridized eigenstates $|\pm\rangle$, rather than from the \emph{nominally mixed} perturbative states $|\mathrm{X_{b(i)}}\rangle^{\rm mix}$ [Fig.~\ref{FIG. 1.}(g)].
Particularly, for $\varepsilon = 3.2\%$, the emission  primarily originates from  $|+\rangle$ state, which is dominated by the $\mathrm{X_b}$ component (similar to the low-strain case), in stead of by the $\mathrm{X_i}$ component.  Below we analyze the underlying physics.

At small strain, $\mathrm{X_b}$ lies substantially below $\mathrm{X_i}$ [Fig.~\ref{FIG. 1.}(d)] owing to the much lower VB maximum at $\Gamma$ valley relative to K valley. Consequently, $|-\rangle$ and $|+\rangle$ are almost purely $\mathrm{X_b}$- and $\mathrm{X_i}$-like, respectively [Regime I in Fig.~\ref{FIG. 1.}(c)],
and only the $\mathrm{X_b}$-dominated  $|-\rangle$  branch emits efficiently, giving rise to the observed single-peak PL at $\varepsilon < 1.6\%$,
in full agreement with experiment [Fig.~\ref{FIG. 1.}(e)].
This scenario behaves as a conventional direct-exciton emitter, with the indirect component being energetically inaccessible for radiative recombination.

As strain increases, the energy detuning between $\mathrm{X_b}$ and $\mathrm{X_i}$ shrinks, strengthening their coupling and causing the $|+\rangle$ branch to acquire increasing $\mathrm{X_b}$ content [cf. Figs.~\ref{FIG. 1.}(c) and \ref{FIG. 1.}(d)]. Once the strain approaches $\varepsilon \simeq 2.0\%$, the two types of bare excitons become nearly degenerate, and the system enters the strongly hybridized scenario [Regime II in Fig.~\ref{FIG. 1.}(c)], giving rise to  pronounced avoided crossing. 
 Note that, across this anticrossing, the dominant characters of $|-\rangle$ and $|+\rangle$ interchange. 
This explains why, at $\varepsilon = 2.4\%$, both emission branches appear with comparable intensity: the $|+\rangle$ state has gained substantial direct-exciton weight, while $|-\rangle$ has begun to lose it. The small energy splitting (only a few meV) further equalizes their thermal populations, so that the $|+\rangle$ state has slight stronger PL intensity than  $|-\rangle$ state.  

At even higher strain [Regime III], e.g.,  $\varepsilon = 3.2\%$, the $|+\rangle$ state dominates the PL emission despite belonging to the higher-energy branch, since it is  is predominantly $\mathrm{X_b}$-derived. Thus, the observed PL does not come from the nominal perturbative state $|\mathrm{X_i}\rangle^{\rm mix}$---which would be $\mathrm{X_i}$-like---but from the genuinely hybridized eigenstate whose character continuously evolves across the avoided crossing. The shaded (gray) region ($\varepsilon = 1.55$–$2.55\%$) in Fig.~\ref{FIG. 1.}(g) marks the strain range where the hybridization is sufficiently strong that the nominally mixed states lose physical meaning.

Figure~\ref{FIG. 1.}(f) shows that the intermediate exciton $\mathrm{X_m}$ plays a crucial role in shaping the PL distribution.
Without $\mathrm{X_m}$ in the relaxation channels,  the $|-\rangle$ branch would dominate the PL at all strain values, in clear contradiction with experiment. Including $\mathrm{X_m}$ introduces a relaxation bottleneck that selectively blocks population flow from $|+\rangle$ to $|-\rangle$, thereby stabilizing the occupation of the upper hybridized state in the relevant strain window.
This routing effect of $\mathrm{X_m}$ is essential for reproducing both the enhancement of PL of the  $|+\rangle$ state  at high strain and the correct strain evolution of the two-peak PL structure.

\paragraph*{``Sinusoidal'' valley control.---}Beyond reproducing the measured PL spectra, our model reveals a striking and highly strain-sensitive valley polarization (VP) response of the hybridized excitons [Figs.~\ref{FIG. 3.}(a) and \ref{FIG. 3.}(b)], demonstrating that strain acts not merely as an energy shifter but as a powerful and versatile control knob for valley physics.

For the upper branch $|+\rangle$ [Fig.~\ref{FIG. 3.}(a)], the VP shows a pronounced nonmonotonic ``sinusoidal''-like evolution that encodes the interplay among direct–indirect hybridization, intervalley pathways, and the intermediate state $\mathrm{X_m}$.
Specifically, at small strain, $|+\rangle$ is predominantly $\mathrm{X_i}$-like, and intravalley relaxation through channel~7 dominates. 
Because the intermediate state $\mathrm{X_m}$ lies above $|+\rangle$ in both valleys---but with a larger detuning in K than in K$^\prime$---the resulting imbalance favors  excitonic population in the K valley, yields a \emph{positive} VP that increases with strain.
Around $\varepsilon\!\sim\!1.3\%$, $\mathrm{X_m^{\rm \Gamma K'}}$ becomes nearly resonant with $|+\rangle$ [Fig.~\ref{FIG. 3.}(c)], greatly enhancing  the direct-indirect mixing.  Then, the intervalley scattering through channel~3  takes over the dominance  and preferentially transfers $\mathrm{X_b}$ population the K valley  to $\mathrm{X_m}$ of the K$^\prime$ valley, suppressing  the VP and even driving it negative for $1.3\%\!\lesssim\!\varepsilon\!\lesssim\!1.7\%$.

At even larger strain, the $|+\rangle$ state rises above both $\mathrm{X_m^{\rm \Gamma K}}$ and $\mathrm{X_m^{\rm \Gamma K'}}$[Fig.~\ref{FIG. 3.}(c)],
activating scattering channel~4 and restoring an increasing VP trend with strain.
The resulting {``sinusoidal''}-like VP profile, with its reversible sign switch, constitutes a continuously tunable \emph{valley dial}, offering deterministic, analog-like control over the valley degree of freedom, far exceeds binary (on/off) control schemes.  Importantly, we verify that this {``sinusoidal''} pattern is basically  resilient against both thermal (temperature) [Fig.~\ref{FIG. 3.}(a)] and magnetic [See the SM; Fig.~S2] effects, underscoring its suitability for realistic device operation.
We emphasize that  this mechanism is  fundamentally distinct from the dark-exciton---assisted biexciton VP reversal in Ref.~\cite{nagler2018}:
here the effect arises from coherent direct-indirect excitonic hybridization mediated by  an intermediate state $\mathrm{X_m}$, not from higher-order excitonic complexes.

The lower branch $|-\rangle$ [Fig.~\ref{FIG. 3.}(b)] shows a markedly different strain evolution, underscoring the selective and \emph{state-specific} nature of strain control. At small strain, $|-\rangle$ is largely $\mathrm{X_b}$-derived;  channel~2 governs relaxation, while  channels~5 and~6 are suppressed,  because $\mathrm{X_m}$ lies far above $|-\rangle$  [Fig.~\ref{FIG. 3.}(d)] and the valley splitting is sizable.
Consequently, the VP is nearly strain-insensitive for $\varepsilon\!\lesssim\!1\%$.
As strain exceeds $\sim 1\%$, the valley splitting of $|-\rangle$ diminishes, enabling enhanced intervalley transfer and reducing VP up to $\varepsilon\!\sim\!2\%$.
For  $2\%\!<\!\varepsilon\!<\!2.7\%$, the EI-assisted intervalley transfer through channel~6 partially restores the VP. 
At even larger strain ($\varepsilon\!\gtrsim\!2.7\%$), channel~5 becomes dominant, strengthening intervalley transfer and driving the VP downward again.
The contrasted responses of the upper and lower hybridized branches demonstrate that strain can potentially selectively address individual excitonic eigenstates, enabling level-resolved selective valley control. 
\paragraph*{Concluding remarks.---}
Our work resolves the fundamental obstacle in understanding momentum-indirect $\rm \Gamma-K$ excitons and establishes the first unified framework that quantitatively reproduces strain-tunable PL in experiments. By revealing the true PL-active hybridized eigenstates, identifying a previously unrecognized  intermediate exciton, and constructing a two-step direct–indirect conversion mechanism, we clarify the microscopic origin of indirect PL. Notably,  we propose a new  concept of
strain-driven ``sinusoidal''  valley control, providing a continuously tunable and highly sensitive \emph{valley dial} far beyond binary switching.
Together with the nearly opposite magnetic-field responses of the two hybridized branches (SM; Sec.~IV), our results  provide a robust
and practical framework for accessing and manipulating long-lived indirect excitons, and open  new avenues toward programmable valley pseudospin engineering and
next-generation valleytronic quantum technologies. As a final remark, our work also offers concrete predictions for forthcoming experiments---especially in light of the very recent strain-controlled VP observations for indirect $\rm Q-K$ excitons~\cite{StrainQ}.

\label{sec:summary}
\paragraph*{Acknowledgments.---} We would like to thank  Paulo Eduardo de Faria Junior and  Zhichu Chen for  valuable comments.
We also wish to acknowledge the late Fanyao Qu, whose generosity in sharing his BSE calculations and whose scientific spirit continue to be warmly remembered.
This work was supported by the National Natural Science Foundation of China (Grants No.~12274256, No.~11874236, No.~12022413, No.~11674331, and No.~61674096), the Major Basic Program of Natural Science Foundation of Shandong Province (Grant No.~ZR2021ZD01),  the National Key R\&D Program of China (Grant No.~2022YFA1403200), and  the ``Strategic Priority Research Program (B)'' of the Chinese Academy of Sciences (Grant No.~XDB33030100).


\begin{thebibliography}{37}
\bibitem [{foo()}]{footnote-exchange}%
  \BibitemOpen
  \href@noop {} {}\bibinfo {note} {Strictly speaking, the momentum mismatch
  between the $\Gamma$ and K valleys endows the intermediate exciton with a
  finite center-of-mass momentum, so the first conversion step may rely on
  phonon-assisted exchange scattering~\cite{darkexciton1}.}\BibitemShut {Stop}%
\end{thebibliography}

%

\end{document}